\documentclass{emulateapj}

\slugcomment{Accepted for publication in ApJ Letters}

\shorttitle{NO GIANT PLANET ORBITING VB\,10}

\shortauthors{Bean et al.}

\begin{document}

\title{The CRIRES Search for Planets Around the Lowest-Mass Stars. \\
II.~The Proposed Giant Planet Orbiting VB\,10 Does Not Exist}

\author{
Jacob L.~Bean\altaffilmark{2,7},
Andreas Seifahrt\altaffilmark{2,3},
Henrik Hartman\altaffilmark{4},
Hampus Nilsson\altaffilmark{4},
Ansgar Reiners\altaffilmark{2,8},
Stefan Dreizler\altaffilmark{2},
Todd J.~Henry\altaffilmark{5},
\& G\"unter Wiedemann\altaffilmark{6},
}

\email{bean@astro.physik.uni-goettingen.de}

\altaffiltext{1}{Based on observations made with ESO Telescopes at the Paranal Observatories under program ID 182.C-0748}

\altaffiltext{2}{Institut f\"ur Astrophysik, Georg-August-Universit\"at,
  Friedrich-Hund-Platz 1, 37077 G\"ottingen, Germany}

\altaffiltext{3}{Department of Physics, University of California, One Shields Avenue, Davis, CA 95616, USA}

\altaffiltext{4}{Lund Observatory, Lund University, Box 43, 22100 Lund, Sweden}

\altaffiltext{5}{Department of Physics and Astronomy, Georgia State University, Atlanta, GA, 30302, USA}

\altaffiltext{6}{Hamburger Sternwarte, Gojenbergsweg 112, 21029 Hamburg, Germany}

\altaffiltext{7}{Marie Curie International Incoming Fellow}

\altaffiltext{8}{Emmy Noether Fellow}

\begin{abstract}
We present high-precision relative radial velocities of the very low-mass star VB\,10 that were obtained over a time span of 0.61\,yr as part of an ongoing search for planets around stars at the end of the main sequence. The radial velocities were measured from high-resolution near-infrared spectra obtained using the CRIRES instrument on the VLT with an ammonia gas cell. The typical internal precision of the measurements is 10\,m\,s$^{-1}$. These data do not exhibit significant variability and are essentially constant at a level consistent with the measurement uncertainties. Therefore, we do not detect the radial velocity variations of VB\,10 expected due to the presence of an orbiting giant planet similar to that recently proposed by Pravdo and Shaklan based on apparent astrometric perturbations. In addition, we do not confirm the $\sim$\,1\,km\,s$^{-1}$ radial velocity variability of the star tentatively detected by Zapatero Osorio and colleagues with lower precision measurements. Our measurements rule out planets with M$_{p}$\,$>$\,3\,M$_{\mathrm{Jup}}$ and the orbital period and inclination suggested by Pravdo and Shaklan at better than 5\,$\sigma$ confidence. We conclude that the planet detection claimed by Pravdo and Shaklan is spurious on the basis of this result. Although the outcome of this work is a non-detection, it illustrates the potential of using ammonia cell radial velocities to detect planets around very low-mass stars.
\end{abstract}

\keywords{planetary systems --- stars: individual (GJ752B)}

\section{INTRODUCTION}
Very low-mass stars are an interesting sample of potential planet hosts. They are useful for probing the correlation between gas giant planet frequency and stellar mass \citep{endl06, johnson07}, are the most numerous stars in the Galaxy \citep{henry06}, and exhibit a larger dynamical response to orbiting planets and have closer-in habitable zones than higher-mass stars \citep{kasting93}. However, the search for planets around these stars is hindered by their extreme intrinsic faintness at visible wavelengths, and few sensitive searches for planets around them have been carried out. 

Recently, \citet[][hereafter PS09]{pravdo09} have claimed the first detection of a planet around a nearby, very low-mass star based on apparent periodic astrometric perturbations. The star is VB\,10 and the planet candidate is a giant planet with a mass M$_{p}$\,=\,6.4\,M$_{\mathrm{Jup}}$ and orbital period P\,=\,0.744\,yr. This could also be the first astrometric discovery of an exoplanet.

We have been observing VB\,10 as part of an ongoing search for planets around the lowest-mass stars (i.e. M$_{\star}$\,$<$\,0.2\,M$_{\odot}$) using high-precision near-infrared (NIR) radial velocity measurements \citep{bean09}. The advantages of the NIR for radial velocity measurements are that it is possible to obtain high-resolution and high-signal-to-noise spectra of very low-mass stars in these wavelengths with current instruments, and also that the affect of activity induced ``jitter'' is potentially reduced relative to the visible \citep{reiners09b}. We present here our radial velocity measurements of VB\,10 obtained so far, and discuss them in the light of PS09's claimed detection. \citet[][hereafter Z09]{zapatero09} have presented NIR radial velocities with a typical internal precision of 300\,m\,s$^{-1}$ that were obtained using NIRSPEC on Keck \citep{mclean98}. These data exhibit variability on the level of $\sim$\,1\,km\,s$^{-1}$, which would be consistent with the existence of a giant planet. Our radial velocity measurements, which are 30 times more precise than those in the Z09 study, also allow us to make a direct comparison to their results.

\section{RADIAL VELOCITY MEASUREMENTS}

\begin{deluxetable}{crrr}
\tabletypesize{\scriptsize}
\tablecolumns{4}
\tablewidth{0pc}
\tablecaption{Radial Velocities Of VB\,10}
\tablehead{
 \colhead{HJD - 2450000.0} &
 \multicolumn{3}{c}{RV (m s$^{-1}$)} 
}
\startdata
4919.89386  &   8.7 & $\pm$ &  7.1\\
4922.92002  &   7.5 & $\pm$ & 15.8\\
4996.79732  & -15.4 & $\pm$ &  5.8\\
4998.78129  &   3.8 & $\pm$ & 10.4\\
4999.75673  & -12.7 & $\pm$ &  6.6\\
5000.78619  &  -3.1 & $\pm$ &  7.2\\
5053.67793  &  13.3 & $\pm$ & 11.8\\
5054.66721  &  19.9 & $\pm$ &  7.5\\
5141.49203  &   2.2 & $\pm$ & 12.5\\
5142.49130  &   7.6 & $\pm$ & 15.5\\
5143.48948  &  11.8 & $\pm$ & 10.0\\
5144.50118  &  -5.1 & $\pm$ & 11.1\\
\enddata
\label{tab:rv}
\end{deluxetable}

We obtained spectroscopic observations of VB\,10 for radial velocity measurements using the high-resolution NIR spectrograph CRIRES \citep{kaeufl04}, which is fed by the UT1 telescope of the Very Large Telescope (VLT) facility. As with all the measurements obtained as part of our CRIRES planet search, a glass cell filled with ammonia gas was placed in front of the spectrograph entrance to enable high-precision calibration of the instrument response during analysis of the data. We used a 0\farcs4 slit for the VB\,10 observations, which corresponds to a nominal resolving power $R  \equiv \lambda / \Delta \lambda \approx$ 50\,000, and we recorded spectra in the CO bandhead region of the $K$-band ($\sim$\,2.3\,$\mu$m). This is a region in which late-type stars exhibit useful lines for radial velocity measurements and the ammonia cell also exhibits useful calibration lines. A total of 36.4\,nm of the 48.2\,nm of recorded wavelength coverage was utilized for the radial velocity measurements.

Observations were obtained during 12 nights in four classical style runs between March and November 2009. Each visit consisted of a sequence (typically four) of 300\,s exposures. We reduced the data and measured relative radial velocities from the extracted spectra using the methods described by us in a previous paper \citep{bean09}. Our radial velocity measurement algorithm is an adaptation of the canonical gas cell technique described by \citet{butler96}, and includes simultaneous modeling of the wavelength scale and instrument point spread function in each spectrum using the ammonia cell lines as a fiducial. We have previously demonstrated that our method can deliver long-term precisions of 5\,m\,s$^{-1}$ for late M dwarfs when applied to data of sufficient quality (i.e. signal-to-noise and resolution) obtained with CRIRES using the ammonia cell. We concentrated only on measuring and analyzing relative radial velocities for this study because our gas cell method enables us to reach velocity precisions on an internal scale that are roughly an order of magnitude better than we could obtain for velocities on an absolute scale. The higher precision we obtain for relative as opposed to absolute velocities yields the best sensitivity for the detection of unseen companions orbiting VB\,10.

The signal-to-noise ratios (SNR) of the reduced and extracted 1d spectra of VB\,10 are $\sim$\,100\,pixel$^{-1}$. The expected photon-limited radial velocity precisions from these data are typically 13\,m\,s$^{-1}$ including calibration uncertainty, and the median of our estimated uncertainties on the radial velocity measurements is 22\,m\,s$^{-1}$. Our estimated uncertainties are normally larger than the corresponding photon-limited precisions because our error estimation method also accounts for the imperfect spectrum modeling during the measurement process. Previous observations of radial velocity standard stars have shown that our adopted measurement uncertainties are reasonable (down to the long-term noise floor of $\sim$\,5\,m\,s$^{-1}$), and we have found that the determined uncertainties for our method are usually a factor of 1.5 -- 2.5 larger than the photon-limited precisions \citep{bean09}. That our estimated uncertainties on the velocities in this case are typically 1.7 times the expected photon-limited error is completely consistent with this previous result. Ultimately, our VB\,10 radial velocities are primarily limited by the obtained SNR in both the spectra that were analyzed to give the measurements, and also the template spectrum utilized for the analysis.

\begin{figure}
\resizebox{\hsize}{!}{\includegraphics{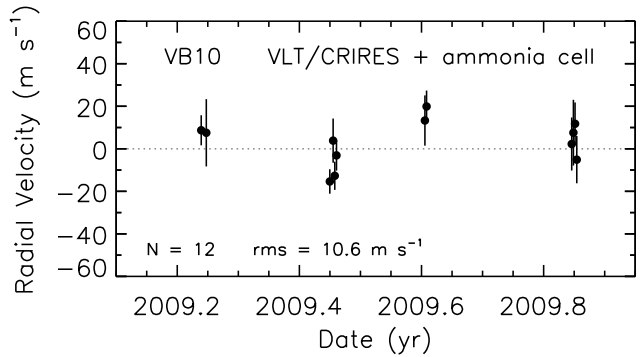}}
\caption{Radial velocities for VB\,10 measured using CRIRES at the VLT with an ammonia gas cell. Multiple exposures taken within up to 2.5\,h have been binned to give each point.}
\label{fig:values}
\end{figure}

We normally combine the radial velocity measurements from spectra taken within a short time frame to give one velocity value. For VB\,10, we binned all observations obtained over 2.5\,hr intervals (4 -- 16 individual measurements) to give the final data set used in this study because this is a short amount of time relative to the orbital period of the putative planet we aimed to detect. The resulting radial velocities are given in Table~\ref{tab:rv} (with the best fit offset subtracted - see the next section) and plotted in Figure \ref{fig:values}. The typical uncertainty in the values is 10\,m\,s$^{-1}$.

\section{ANALYSIS}
\subsection{Comparison of models}
The radial velocity measurements we have obtained exhibit an rms dispersion of 10.6\,m\,s$^{-1}$ around a constant value when binned over 2.5\,h, which is consistent with our estimated uncertainties. Therefore, we have not detected reflex orbital motion induced on VB\,10 by an unseen companion. Therefore, the question becomes whether or not our data are inconsistent with the prediction of the existence of a giant planet from PS09, and the potential detection of radial velocity variability by Z09.

\begin{figure}
\resizebox{\hsize}{!}{\includegraphics{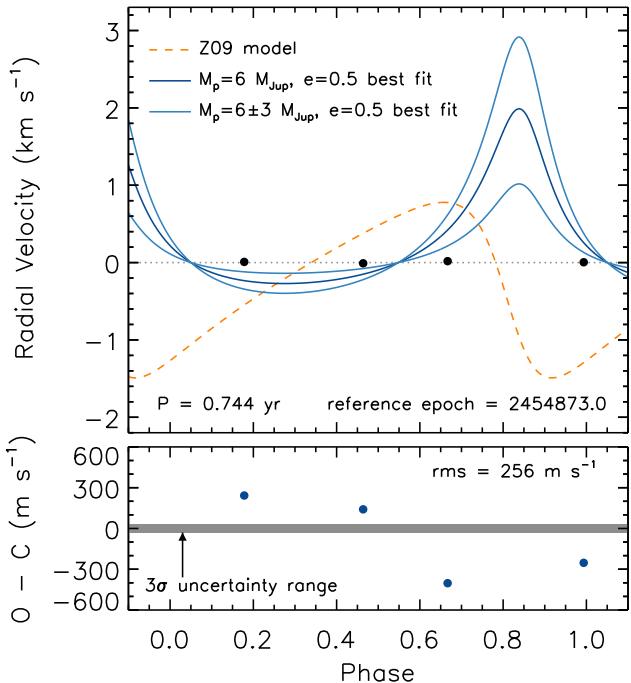}}
\caption{\textit{Top} Measured radial velocities for VB\,10 phased to the orbital period given by PS09 (points). The data are binned over each observing run for clarity and the error bars are smaller than the symbols. The dashed orange line gives the orbit model suggested by Z09 with the relative offset determined to give the best fit. The solid dark blue line gives the best fit to the data for a model with a 6\,M$_{\mathrm{Jup}}$ planet on an orbit with e\,=\,0.5. The light blue solid lines give the best fit models for 3\,M$_{\mathrm{Jup}}$ (smaller semi-amplitude) and 9\,M$_{\mathrm{Jup}}$ (larger semi-amplitude) planets also with e=0.5. \textit{Bottom} Residuals from the best fit 6\,M$_{\mathrm{Jup}}$ planet model (points). The grey bar gives the range of values that would be consistent with three times the observed dispersion in the data ($\pm$\,32\,m\,s$^{-1}$).}
\label{fig:compare}
\end{figure}

Our measured radial velocities are plotted along with some possible orbit models in Figure \ref{fig:compare}. The data span 82\% of the orbital period (0.744\,yr) suggested by PS09, in steps of $\sim$\,0.25 phase units. Given that our uncertainties are roughly two orders of magnitude smaller than the semi-amplitude expected for a giant planet with an orbital period of a few hundred days ($\gtrsim$\,1\,km\,s$^{-1}$), it is likely that we would have detected significant changes in the star's radial velocity over the time span the data were obtained if such a planet exists. 

As discussed in \S2, our measurements, like all other high-precision radial velocity measurements, are relative to an arbitrary scale. We have determined a best fit constant offset for all the orbits shown in this paper to account for this. We have fixed the orbital period and inclination ($i$\,=\,96.9$\degr$) to the values given by PS09 for all the analyses presented in this paper because our aim is to assess their specific claim of a planet detection. The corresponding uncertainties in the period and inclination given by PS09 do not have a significant influence on the models relative to the precision of our data. For example, adjusting the assumed period and inclination within their 1\,$\sigma$ uncertainties to obtain the smallest possible velocity semi-amplitude can only yield a maximum reduction in the semi-amplitude of $<$\,3\%. In addition to the period and inclination, we also assumed the same mass for VB\,10 that PS09 adopted (M$_{\star}$\,=\,0.078\,M$_{\odot}$) to stay consistent with their study.

The orbit model that was found by Z09 to give the best fit to both their NIRSPEC radial velocities and the PS09 astrometry is shown in Figure \ref{fig:compare} for comparison with our data. The rms of the residuals of our data from this model with the relative offset optimized is 870\,m\,s$^{-1}$. Therefore, the Z09 model is inconsistent with our measurements. 

Some additional orbit models for planets with masses of 3, 6, and 9\,M$_{\mathrm{Jup}}$ are shown in Figure \ref{fig:compare}. This range of masses roughly corresponds to the nominal value and upper and lower 1\,$\sigma$ confidence limits determined by PS09. The time and location of periastron, and the eccentricity of the candidate planet detected by PS09 are not constrained by their astrometry data. For these example models we adopted a fixed eccentricity e\,=\,0.5, and optimized the remaining parameters to give the best fit to the data (time of periastron, longitude of periastron, and radial velocity offset). As can be seen in the figure, the optimization of the unconstrained parameters leads to the orbital phases when the largest velocity variations are expected to be outside the times covered by our observations. For reference, the best fit time of periastron is 2455103, and the best fit longitude of periastron is 10$\degr$. However, the models do not fit the data even with these optimizations. The lower panel in Figure \ref{fig:compare} shows the residuals from the M$_{p}$\,=\,6\,M$_{\mathrm{Jup}}$ model. The rms of the residuals is 256\,m\,s$^{-1}$, which is much larger than the observed dispersion in the data and the uncertainties in the individual measurements.

\subsection{Limits to a giant planet}
We conclude from the comparisons described above that it is unlikely that our radial velocity measurements are compatible with the presence of a giant planet around VB\,10 with an orbital period of 0.744\,yr and nearly edge-on orientation. Therefore, we were motivated to undertake a more specific and robust determination of what the data rule out. We did this by mapping the quality of orbit model fits to the data as a function of the possible parameters, and comparing the determined values to the fit quality of a model representing no radial velocity variations (i.e. a flat line). It can be seen from inspection of Figure \ref{fig:compare} that higher eccentricity and lower mass planets could be more easily hidden in the data. This led us to map the fit quality as a function of possible planet mass and orbital eccentricity. We considered planet masses ranging from 0.5 to 15\,M$_{\mathrm{Jup}}$, and eccentricity values ranging from 0.0 to 0.95. We again fixed the period and inclination to those values given by PS09, and we marginalized over the unconstrained parameters at each grid point.

\begin{figure}
\resizebox{\hsize}{!}{\includegraphics{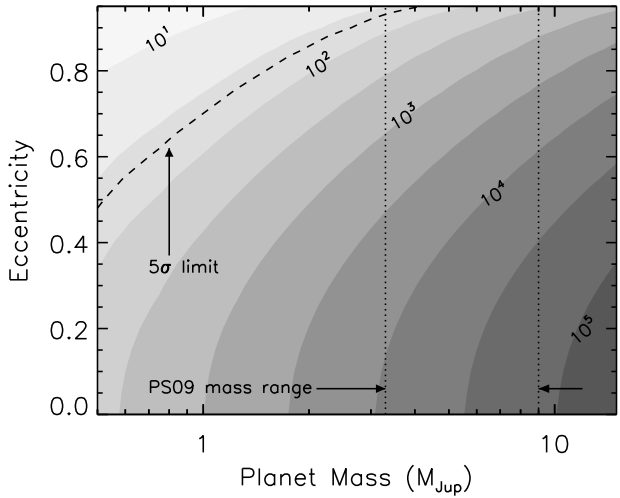}}
\caption{Map of the best fit $\chi^{2}$ for a planet with P\,=\,0.744\,yr and $i$\,=\,96.9$\degr$. At each point in the map the time of periastron, longitude of periastron, and radial velocity offset were optimized to give the best fit. The 5\,$\sigma$ upper limit is given by the dashed contour. The region to the lower right of this line is excluded by our data. The dotted lines delineate the 1\,$\sigma$ confidence interval on the planet mass from PS09.}
\label{fig:map}
\end{figure}

Figure \ref{fig:map} shows the obtained map of the fit $\chi^{2}$. For reference, the $\chi^{2}$ for a fit of a flat line to the data is 23.1 with 11 degrees of freedom. This somewhat higher than expected $\chi^{2}$ suggests that either VB\,10 is exhibiting real radial velocity variability at the 10\,m\,s$^{-1}$ level, jitter arising from activity is affecting the measurements also at about the 10\,m\,s$^{-1}$ level \citep[note that VB\,10 is known to be a variable source with some flare activity,][]{berger08}, we have underestimated our uncertainties by about 50\%, or some combination of these effects. At this point it is not critical to understand the origin of the additional variability in the radial velocity measurements because this investigation is focused on ruling out much larger variations.

The 5\,$\sigma$ limit shown in Figure \ref{fig:map} corresponds to the boundary where the fit $\chi^{2}$ increases by more than 25 from the $\chi^{2}$ of the flat line fit to the data. Everything to the lower right of this line is ruled out at high confidence. We find that for all conceivable eccentricity values (i.e. e\,$<$\,0.95) we can rule out planets with M$_{p}$\,$>$\,3\,M$_{\mathrm{Jup}}$ and the orbital period and inclination suggested by PS09. This means that we can definitively rule out planets with masses in PS09's given 1\,$\sigma$ confidence interval (indicated by the dotted lines in Figure \ref{fig:map}). All higher mass planets are also ruled out. At the lower range of possible planet masses the determined limit is dependent on the assumed eccentricity. For example, planet masses of 1\,M$_{\mathrm{Jup}}$ and 0.5\,M$_{\mathrm{Jup}}$ can only be ruled out for e\,$<$0.7 and e\,$<$\,0.5 respectively.

\section{DISCUSSION}
We have presented radial velocities for VB\,10 that were obtained in the NIR using a new type of gas cell and that have a typical internal precision of 10\,m\,s$^{-1}$. These data were obtained in four epochs over a time span of 0.61\,yr. Our radial velocity measurements do not exhibit significant variability and are essentially constant at a level consistent with the measurement uncertainties. Therefore, we see no evidence for a giant planet orbiting VB\,10 similar to that detected via apparent astrometric perturbations by PS09. In addition, we do not confirm the $\sim$\,1\,km\,s$^{-1}$ variability tentatively detected by Z09 using NIRSPEC radial velocity measurements that have a typical internal precision of 300\,m\,s$^{-1}$. Our data are also inconsistent with the orbit model presented by Z09 as giving the best fit to both the PS09 astrometry and their radial velocities combined. If we consider our radial velocity data alone, then we can rule out planets with M$_{p}$\,$>$\,3\,M$_{\mathrm{Jup}}$ in the orbit suggested by PS09. Planets with masses down to 1\,M$_{\mathrm{Jup}}$ would need to have unusually large orbital eccentricities and be phased in a particular way to not have caused highly significant variations in our measured radial velocities. 

After we submitted our paper we became aware of another paper discussing the proposed VB\,10 planet that was submitted at roughly the same time \citep{anglada-escude10}. The authors of that paper presented radial velocity measurements for VB10 with a typical precision of $\sim$\,200\,m\,s$^{-1}$ and that covered roughly the same time period as our measurements. Their data also indicate a non-detection of the reflex motion of VB\,10 expected due to the presence of the giant planet proposed by PS09. \citet{anglada-escude10} carried out a combined analysis of the PS09 astrometry and their radial velocities and found that the possible orbit models do not give a significantly better fit to the data than a model assuming no reflex motion. They concluded that the giant planet proposed by PS09 is not likely to exist on the basis of these results, although they couldn't completely rule it out with their data.

Our results seem to completely rule out the presence of a giant planet similar to that proposed by PS09, which is consistent with the results of \citet{anglada-escude10}. Therefore, we suggest that the planet detection claimed by PS09 is spurious. If further work verifies this result then it would still be the case that no planet has yet been discovered via astrometry, and that the current state-of-the-art for the astrometric study of exoplanets remains post priori detection and characterization of a few select planets that were first identified with the radial velocity method \citep[e.g.][]{benedict02, mcarthur04, bean07, martioli09}. We note that obtaining astrometric measurements at the level of precision needed to detect exoplanets is notoriously difficult, especially from the ground as PS09 have attempted to do. There is a long history of claimed and later refuted astrometric exoplanet detections dating back to the famous case of the possible planet around Barnard's star, which was originally suggested based on 25 years of observations by \citet{vandekamp63} but later refuted by \citet{hershey73}, \citet{gatewood73}, and others. Furthermore, PS09 were not even the first to propose the existence of a sub-stellar companion to VB\,10 based on apparent astrometric perturbations \citep{harrington83}. The first candidate was later disproved by subsequent astrometric measurements \citep{monet92,harrington93}.

With regards to the possible radial velocity variations in VB\,10 seen by Z09, we suggest their analysis could have been influenced by an important systematic effect. This is a possibility that those authors themselves acknowledged. Of the five measurements presented by Z09, the one point in their data set that is inconsistent with a constant radial velocity (i.e. no variations possibly due to orbital motion) is also based on a spectrum that was taken with a different slit width than all the other data. Z09 did not use a deconvolution technique to account for changes in the spectrograph point spread function, so the velocity measured from this spectrum could have a different relative zero point compared to the other data. It is not clear whether the different slit width would cause as large a discrepancy as they saw ($\sim$1\,km\,s$^{-1}$), but it is a possible candidate for reconciling the difference between their measurements and ours. 

Despite the negative result indicated by our data, we propose that there is a positive general result from this work. Namely, that we have demonstrated the power of NIR radial velocities obtained using CRIRES at the VLT with our ammonia cell. Late-type M dwarfs are an interesting potential planet host sample, but they are also extremely challenging for the normal planet search methods due to their faintness. For example, VB\,10 is one of the closest, and therefore brightest, stars with an estimated mass $<$\,0.1\,M$_{\odot}$, but it still has $V$\,=\,17.5. This puts it far beyond the reach of high-precision radial velocity measurements in the visible with current instruments. However, late M dwarfs are reasonably bright in the NIR due to their extreme redness, and VB\,10 has $K$\,=\,8.8. By developing a technique for measuring high-precision NIR radial velocities, we have opened the door to a sensitive search for planets around stars at the very end of the main sequence and even beyond. The method is a valuable stand-alone tool, and also a complement to the ongoing astrometric planet searches targeting M dwarfs \citep[e.g.][]{pravdo04, henry06, boss09, forbrich09}. We are continuing to obtain radial velocity measurements for VB\,10, and many other very low-mass stars to search for planets. The results from this work will help constrain the nature of planets around the most numerous stars in the Galaxy.

\acknowledgments
We thank the ESO staff for their assistance with our program. This work was partly supported by the DFG through grants GRK 1351 and RE 1664/4-1, and the BMBF through program 05A0GU2. J.L.B. acknowledges research funding from the European Commission’s Seventh Framework Program as an International Fellow (grant no.~PIFF-GA-2009-234866). A.S. acknowledges financial support from the NSF under grant AST-0708074. H.H. acknowledges funding from the Swedish Research Council (VR). H.N. acknowledges the financial support from the Lund Laser Center through a Linnaeus grant from the Swedish Research Council (VR). A.R. received support from the DFG as an Emmy Noether Fellow.

{\it Facilities:} \facility{VLT:Antu (CRIRES)}

\bibliographystyle{apj}
\bibliography{ms.bib}

\begin{thebibliography}{26}
\expandafter\ifx\csname natexlab\endcsname\relax\def\natexlab#1{#1}\fi

\bibitem[{{Anglada-Escude} {et~al.}(2010){Anglada-Escude}, {Shkolnik},
  {Weinberger}, {Thompson}, {Osip}, \& {Debes}}]{anglada-escude10}
{Anglada-Escude}, G., {Shkolnik}, E.~L., {Weinberger}, A.~J., {Thompson},
  I.~B., {Osip}, D.~J., \& {Debes}, J.~H. 2010, ApJL submitted, arXiv:1001.0043

\bibitem[{{Bean} {et~al.}(2007){Bean}, {McArthur}, {Benedict}, {Harrison},
  {Bizyaev}, {Nelan}, \& {Smith}}]{bean07}
{Bean}, J.~L., {McArthur}, B.~E., {Benedict}, G.~F., {Harrison}, T.~E.,
  {Bizyaev}, D., {Nelan}, E., \& {Smith}, V.~V. 2007, \aj, 134, 749

\bibitem[{{Bean} {et~al.}(2009){Bean}, {Seifahrt}, {Hartman}, {Nilsson},
  {Wiedemann}, {Reiners}, {Dreizler}, \& {Henry}}]{bean09}
{Bean}, J.~L., {Seifahrt}, A., {Hartman}, H., {Nilsson}, H., {Wiedemann}, G.,
  {Reiners}, A., {Dreizler}, S., \& {Henry}, T.~J. 2009, \apj\ submitted,
  arXiv:0911.3148

\bibitem[{{Benedict} {et~al.}(2002){Benedict}, {McArthur}, {Forveille},
  {Delfosse}, {Nelan}, {Butler}, {Spiesman}, {Marcy}, {Goldman}, {Perrier},
  {Jefferys}, \& {Mayor}}]{benedict02}
{Benedict}, G.~F., {et~al.} 2002, \apjl, 581, L115

\bibitem[{{Berger} {et~al.}(2008){Berger}, {Basri}, {Gizis}, {Giampapa},
  {Rutledge}, {Liebert}, {Mart{\'{\i}}n}, {Fleming}, {Johns-Krull}, {Phan-Bao},
  \& {Sherry}}]{berger08}
{Berger}, E., {et~al.} 2008, \apj, 676, 1307

\bibitem[{{Boss} {et~al.}(2009){Boss}, {Weinberger}, {Anglada-Escud{\'e}},
  {Thompson}, {Burley}, {Birk}, {Pravdo}, {Shaklan}, {Gatewood}, {Majewski}, \&
  {Patterson}}]{boss09}
{Boss}, A.~P., {et~al.} 2009, \pasp, 121, 1218

\bibitem[{{Butler} {et~al.}(1996){Butler}, {Marcy}, {Williams}, {McCarthy},
  {Dosanjh}, \& {Vogt}}]{butler96}
{Butler}, R.~P., {Marcy}, G.~W., {Williams}, E., {McCarthy}, C., {Dosanjh}, P.,
  \& {Vogt}, S.~S. 1996, \pasp, 108, 500

\bibitem[{{Endl} {et~al.}(2006){Endl}, {Cochran}, {K{\"u}rster}, {Paulson},
  {Wittenmyer}, {MacQueen}, \& {Tull}}]{endl06}
{Endl}, M., {Cochran}, W.~D., {K{\"u}rster}, M., {Paulson}, D.~B.,
  {Wittenmyer}, R.~A., {MacQueen}, P.~J., \& {Tull}, R.~G. 2006, \apj, 649, 436

\bibitem[{{Forbrich} \& {Berger}(2009)}]{forbrich09}
{Forbrich}, J., \& {Berger}, E. 2009, \apjl, 706, L205

\bibitem[{{Gatewood} \& {Eichhorn}(1973)}]{gatewood73}
{Gatewood}, G., \& {Eichhorn}, H. 1973, \aj, 78, 769

\bibitem[{{Harrington} {et~al.}(1983){Harrington}, {Kallarakal}, \&
  {Dahn}}]{harrington83}
{Harrington}, R.~S., {Kallarakal}, V.~V., \& {Dahn}, C.~C. 1983, \aj, 88, 1038

\bibitem[{{Harrington} {et~al.}(1993){Harrington}, {Dahn}, {Kallarakal},
  {Guetter}, {Riepe}, {Walker}, {Pier}, {Vrba}, {Luginbuhl}, {Harris}, \&
  {Ables}}]{harrington93}
{Harrington}, R.~S., {et~al.} 1993, \aj, 105, 1571

\bibitem[{{Henry} {et~al.}(2006){Henry}, {Jao}, {Subasavage}, {Beaulieu},
  {Ianna}, {Costa}, \& {M{\'e}ndez}}]{henry06}
{Henry}, T.~J., {Jao}, W., {Subasavage}, J.~P., {Beaulieu}, T.~D., {Ianna},
  P.~A., {Costa}, E., \& {M{\'e}ndez}, R.~A. 2006, \aj, 132, 2360

\bibitem[{{Hershey}(1973)}]{hershey73}
{Hershey}, J.~L. 1973, \aj, 78, 421

\bibitem[{{Johnson} {et~al.}(2007){Johnson}, {Butler}, {Marcy}, {Fischer},
  {Vogt}, {Wright}, \& {Peek}}]{johnson07}
{Johnson}, J.~A., {Butler}, R.~P., {Marcy}, G.~W., {Fischer}, D.~A., {Vogt},
  S.~S., {Wright}, J.~T., \& {Peek}, K.~M.~G. 2007, \apj, 670, 833

\bibitem[{{Kasting} {et~al.}(1993){Kasting}, {Whitmire}, \&
  {Reynolds}}]{kasting93}
{Kasting}, J.~F., {Whitmire}, D.~P., \& {Reynolds}, R.~T. 1993, Icarus, 101,
  108

\bibitem[{{K\"aufl} {et~al.}(2004){K\"aufl}, {Ballester}, {Biereichel},
  {Delabre}, {Donaldson}, {Dorn}, {Fedrigo}, {Finger}, {Fischer}, {Franza},
  {Gojak}, {Huster}, {Jung}, {Lizon}, {Mehrgan}, {Meyer}, {Moorwood}, {Pirard},
  {Paufique}, {Pozna}, {Siebenmorgen}, {Silber}, {Stegmeier}, \&
  {Wegerer}}]{kaeufl04}
{K\"aufl}, H.-U., {et~al.} 2004, in SPIE Conference Series, ed. A.~F.~M.
  {Moorwood} \& M.~{Iye}, Vol. 5492, 1218--1227

\bibitem[{{Martioli} {et~al.}(2010){Martioli}, {McArthur}, {Benedict}, {Bean},
  {Harrison}, \& {Armstrong}}]{martioli09}
{Martioli}, E., {McArthur}, B.~E., {Benedict}, G.~F., {Bean}, J.~L.,
  {Harrison}, T.~E., \& {Armstrong}, A. 2010, \apj, 708, 625

\bibitem[{{McArthur} {et~al.}(2004){McArthur}, {Endl}, {Cochran}, {Benedict},
  {Fischer}, {Marcy}, {Butler}, {Naef}, {Mayor}, {Queloz}, {Udry}, \&
  {Harrison}}]{mcarthur04}
{McArthur}, B.~E., {et~al.} 2004, \apjl, 614, L81

\bibitem[{{McLean} {et~al.}(1998){McLean}, {Becklin}, {Bendiksen}, {Brims},
  {Canfield}, {Figer}, {Graham}, {Hare}, {Lacayanga}, {Larkin}, {Larson},
  {Levenson}, {Magnone}, {Teplitz}, \& {Wong}}]{mclean98}
{McLean}, I.~S., {et~al.} 1998, in SPIE Conference Series, ed. A.~M. {Fowler},
  Vol. 3354, 566--578

\bibitem[{{Monet} {et~al.}(1992){Monet}, {Dahn}, {Vrba}, {Harris}, {Pier},
  {Luginbuhl}, \& {Ables}}]{monet92}
{Monet}, D.~G., {Dahn}, C.~C., {Vrba}, F.~J., {Harris}, H.~C., {Pier}, J.~R.,
  {Luginbuhl}, C.~B., \& {Ables}, H.~D. 1992, \aj, 103, 638

\bibitem[{{Pravdo} \& {Shaklan}(2009)}]{pravdo09}
{Pravdo}, S.~H., \& {Shaklan}, S.~B. 2009, \apj, 700, 623, PS09

\bibitem[{{Pravdo} {et~al.}(2004){Pravdo}, {Shaklan}, {Henry}, \&
  {Benedict}}]{pravdo04}
{Pravdo}, S.~H., {Shaklan}, S.~B., {Henry}, T., \& {Benedict}, G.~F. 2004,
  \apj, 617, 1323

\bibitem[{{Reiners} {et~al.}(2009){Reiners}, {Bean}, {Huber}, {Dreizler},
  {Seifahrt}, \& {Czesla}}]{reiners09b}
{Reiners}, A., {Bean}, J.~L., {Huber}, K.~F., {Dreizler}, S., {Seifahrt}, A.,
  \& {Czesla}, S. 2009, \apj\ submitted, arXiv:0909.0002

\bibitem[{{van de Kamp}(1963)}]{vandekamp63}
{van de Kamp}, P. 1963, \aj, 68, 515

\bibitem[{{Zapatero Osorio} {et~al.}(2009){Zapatero Osorio}, {Mart{\'{\i}}n},
  {del Burgo}, {Deshpande}, {Rodler}, \& {Montgomery}}]{zapatero09}
{Zapatero Osorio}, M.~R., {Mart{\'{\i}}n}, E.~L., {del Burgo}, C., {Deshpande},
  R., {Rodler}, F., \& {Montgomery}, M.~M. 2009, \aap, 505, L5, Z09

\end{thebibliography}

\end{document}